Cyber Behavior of Microblog Users: Onlies Versus Others

Dong Nie, Bibo Hao, Zheng Yan, Tingshao Zhu*

Institute of Psychology, Chinese Academy of Sciences, China

School of Computer and Control Engineering, University of Chinese Academy of Sciences, China

Author Note

The authors gratefully acknowledge the generous support from NSFC (61070115), Institute of Psychology (113000C037), Strategic Priority Research Program (XDA06030800) and 100-Talent Project (Y2CX093006) from Chinese Academy of Sciences.

Abstract

Much research has been conducted to investigate personality and daily behavior of these only children ("Onlies") due to the Chinese one-child-per-family policy, and report the singleton generation to be more selfish. As Microblog becomes increasingly popular recently in China, we studied cyber behavior of Onlies and children with siblings ("Others") on Sina Microblog ("Weibo"), a leading Microblog service provider in China. Participants were 1792 Weibo users. Their recorded data on Weibo were downloaded to assess their cyber behaviors. The general results show that (1) Onlies have a smaller social circle; (2)Onlies are more significantly active on social platform.

Cyber Behavior of Microblog Users: Onlies Versus Others

**Introduction**

More than three decades have passed since the execution of the one-child-per-family policy in China in 1979. Since then, the fertility rate in China fell from 2.63 births per woman in 1980 (already a sharp reduction from more than five births per woman in the early 1970s) to 1.6 in 2007. As of 2007, approximately 35.90% of China's population was subjected to a one-child restriction. Much research (e.g., Jiao & Ji, 1986; Falbo & Poston, 1993) has been done to compare daily behaviors between only children ("Onlies") and children with siblings ("Others"), and to examine the effects of one-child policy on children development. It is usually hypothesized that Onlies are spoiled, egocentric, maladjusted and less cooperative, and they are often called "little emperors". However, there are a lot of controversy observational studies over the hypothesis of Onlies. (Hvistendahl, 2012) reported that people born after the introduction of the one-child policy were not only less trusting, less trustworthy, and more pessimistic, but also less competitive, less conscientious, and more risk-averse. Different from Hvistendahl's finding, other research found that Onlies performed better in several fields such as motivation to achieve success, and always performed as well as Others in daily behavior. The controversy of their daily behaviors motivates us to further this research and to explore the online behaviors between Onlies and Others.

Sina Weibo, which is China's twitter, is now one of the most popular internet services in mainland China, with more than 300 million registered users (Millward,

2012). Many people spend much time on Weibo, and it is said that Weibo has become an important part of human life. Much research has been conducted based on Microblog platforms (Golbeck, 2011), (Gao, Abel, Houben & Yu, 2012) mainly sum up the online behaviors, for example, geographical distribution. Nevertheless, the difference comparison between Onlies and Others has not ever been conducted.

In this paper, we used a behavior analysis approach to look into how Onlies differ from Others on Microblog platform. We acquired 1792 Weibo users' public information using Sina Weibo Application Programming Interface (API). We run various analysis methods to process the data, and we intended to find the differences between Onlies and Others Microblog behaviors. To make the single child effect trustful, we also ran analysis on demographic and background factors to identify the real effect.

The rest of this paper is organized as follows. We describe some related work at first, and then introduce the dataset in detail. Experiment results are presented in Experiment section. The discussions of one-child-per-family policy will be presented in Discussion section, and we conclude in Conclusion section.

## Related Work

There has been much work on assessing the differences between Onlies and Others, especially in China due to one-child policy after 1979. Quite a few research have focused on the same problem but reached opposite conclusions.

Some studies support the assumption that Onlies in China is a spoiled generation. (Jiao & Ji, 1986) enrolled 933 children to take part in their study, and found that Onlies are more egocentric, whereas Others possess the positive qualities of persistence, cooperation, and peer prestige. (Hall, 1987) take similar comparative studies on behaviors and personality as well as mental health, and their conclusions generally support the assumption.

Other researches assume that Onlies and Others are the same. (Falbo & Poston, 1993) conducted a study of 1,000 school children, and found that very few Onlies effects (Onlies were totally different from Others, and they were a spoiled generation) were detected. Even more, Onlies outperformed Others on both academics and physical. (Poston & Falbo, 1990) also suggested that the one-child policy in China was not producing a generation of "little emperors". A few researchers even believed that Onlies had lower levels of fear, anxiety and depression (Yang & Ollendick, 1995). However, most of the above observations are derived from parents, teachers or peers, and self-report measurements are seldom taken to deal with subjects. To the best of our knowledge, this is the first attempt to investigate the difference in social media behaviors between Onlies and Others until now.

Psychological analysis on social media has received considerable attention recently (Qiu & Lin, 2012). Personality has been reported to be relevant to many types of behaviors, and it even can be predicted from social media profile (Golbeck & Robles, 2011). It has been shown in (Rosen & Kluemper, 2008) that extroversion and conscientiousness positively correlate with the perceived use of social media sites. Extroversion has a positive correlation with perceived usage with less long sentences, less complex writings, and more social and positive emotional words (Qiu & Lin,

2012). Agreeable individuals would like to use more positive emotion words and first person plural pronouns (Yarkoni, 2010; Selfhout & Burk, 2010) show that people often make friends with individuals with high agreeableness, and they tend to choose friends with similar agreeableness, extroversion, and openness scores. (Schrammel & Koffel, 2009) find a correlation between openness and the number of friends. In general, people with different personality behave differently in social media sites.

Our research investigates the difference in online behaviors between Onlies and Others, and there are two major contributions. The first is that subjects are in such a large scale and randomly selected almost over the country instead of choosing from the same place. The other is we compare the social media behavior between Onlies and Others for the first time, and social media behaviors are objective rather than subjective.

In this paper, we attempted to address one research questions: Are there any differences in Microblog behaviors between Onlies and Others? Furthermore, we tried to ensure where these differences are caused by single child factor. We collected their Microblog data to measure subjects' behaviors objectively. We then compared the difference between Onlies and Others participators using statistical analysis. At last, we discuss the experiment results from the view of psychological perspective.

## Method

The dataset consists of 1792 Microblog users' usage data. The Microblog data can be spilt into several categories as follows:

1. Personal Presentation: including nickname, personalized domain name,

description and so on.

2. Social Circle: friend, follower and mutual follower.

2. Social Activity: status, repost status, comment, picture,

4. Social Habit: time to post status.

**Data Collection**

Using Sina Weibo API, we first collected about 100M Weibo user IDs, randomly chosen 20,000 user IDs, and crawled their Weibo data. By using Weibo's'@' function, we invited the users to participate our study.  Among them, 933 subjects are Onlies, and the remaining are Others. The average age of Onlies is 23.2, and Others is 23.4. Some statistical results about the subjects are listed as follows in Table 1.

Table 1

*Descriptive Statistics of Subjects*

| Variables | | Population | Percentage |
| --- | --- | --- | --- |
| Hometown location | City | 1087 | 60.66 |
| | Town | 337 | 18.80 |
| | Rural area | 368 | 20.54 |
| Family monthly income(CNY) | <2000 | 277 | 15.46 |
| | 2001-4000 | 648 | 36.16 |
| | 4001-6000 | 338 | 18.86 |
| | 6001-8000 | 219 | 12.22 |
| | 8001-10000 | 136 | 7.59 |
|  | >10000 | 174 | 9.71 |
| Occupation | non-student | 913 | 50.95 |
|  | Student | 879 | 49.05 |
| Age | >32 | 98 | 5.47 |
|  | <=32 | 1694 | 94.53 |
| Gender | Male | 653 | 36.44 |
|  | Female | 1139 | 63.56 |

*Note.* n = 1792.

As shown in Table 1, most of the subjects are birthed after 1979, the subjects' family monthly income generally meet the real income situation of Chinese, and about 40% of subjects come from rural area or town.

The whole process took over one month, and volunteers had got reimbursement in return. The collected Weibo-user dataset (1792copies of labeled Weibo data) was then preprocessed for further analysis.

**Feature Extraction**

As the collected Weibo data is raw data, the first step is to extract behavior features. We totally extract 20 features for one user from the Weibo data. The features are listed in Table 2.

Table 2

*Extracted Features*

| Feature | Description |
|---|---|
| **Personal Presentation** | |
| screen name length | The length of screen name |
| description evaluation | The sentiment evaluation of user's description |
| description length | The length of user's description |
| tag count | The total count of tags |
| **Social Circle** | |
| mutual follower | The number of user's mutual followers |
| follower | The number of user's followers |
| friend | The number of user's friends |
| **Social Activity** | |
| status | The number of user's statuses |
| original status | The number of user's original statuses |
| original status rate | The rate of original statuses |
| repost status | The number of user's reposted statuses |
| favorite status | The number of user's favorite statuses |
| Comment | The number of comments |
| comment variance | The variance of comments per status |
| original picture | The number of users' original pictures |
| picture per status | The average number of pictures per status |
| picture rate | The rate of original picture |
| **Social Habit** | |
| period for 1st status | The period subject most likely to give first status |
| period for last period | The period subject most likely to give last status |
| period for most statuses | The period subject to give most statuses |

Some features are simply from the original data, for example, the number of statuses. We ran a program to analysis users' descriptions, to identify each description as positive, neuter or negative. To calculate the time of creating statuses, we divide a whole day into 7 periods: 0:00-6:00, 6:00-8:00, 8:00-11:00, 11:00-13:00, 13:00-17:00, 17:00-20:00, and 20:00-24:00. We then count the number of statuses that user posts in each period, thus the last three features in Table 2 are extracted.

## Results

**Cyber Behavior Differences between Onlies and Others**

We compare microblog behaviors between two groups through independent t-test method. Most of the group statistics results are listed in Table 3, and the major independent t-test results are followed in Table 4.

Table 3

*Group Statistics and Independent Samples T Test of Microblog Behaviors*

| Microblog behavior | Child Type | $\bar{x}$ | S.D. | t | df | p |
|---|---|---|---|---|---|---|
| **Personal Presentation** | | | | | | |
| screenname length | Onlies | 14.62 | 6.88 | -1.47 | 1790 | .14 |
| | Others | 15.10 | 7.13 | | | |
| description length | Onlies | 51.74 | 47.44 | 0.54 | 1790 | .59 |
| | Others | 50.55 | 46.68 | | | |
| **description evaluation** | Onlies | 2.24 | 0.62 | 2.40 | 1790 | <.05 |
| | Others | 2.17 | 0.65 | | | |
| tag count | Onlies | 4.80 | 3.82 | -1.38 | 1790 | .17 |
| | Others | 4.60 | 3.80 | | | |
| **Social Circle** | | | | | | |
| **mutual follower** | Onlies | 130.65 | 133.24 | -3.79 | 1790 | <.01 |
| | Others | 160.75 | 198.54 | | | |
| **follower** | Onlies | 710.12 | 1379.67 | -.1.67 | 1790 | <.10 |
| | Others | 1105.55 | 7093.86 | | | |
| friend | Onlies | 360.93 | 319.24 | -0.67 | 1790 | .50 |
| | Others | 371.43 | 342.93 | | | |
| **Social Activity** | | | | | | |
| **status** | Onlies | 2834.19 | 1376.10 | 3.02 | 1790 | <.01 |
| | Others | 2642.42 | 1307.04 | | | |
| original status | Onlies | 801.17 | 678.72 | 1.18 | 1790 | .24 |
| | Others | 764.54 | 632.67 | | | |
| original status rate | Onlies | 0.29 | 0.20 | -1.34 | 1790 | .18 |
| | Others | 0.31 | 0.20 | | | |
| repost status | Onlies | 569.51 | 1610.74 | 1.45 | 1790 | .15 |
| | Others | 477.89 | 953.22 | | | |
| favorite status | Onlies | 313.96 | 986.49 | 0.38 | 1790 | .71 |
| | Others | 296.26 | 991.30 | | | |
| Comment | Onlies | 1575.06 | 2135.50 | -.03 | 1790 | .97 |
| | Others | 477.89 | 2001.27 | | | |
| comment variance | Onlies | 97.56 | 98.86 | -1.36 | 1790 | .17 |
| | Others | 105.05 | 133.37 | | | |
| **original picture** | Onlies | 334.15 | 345.60 | 5.70 | 1790 | <.01 |
| | Others | 236.58 | 379.24 | | | |
| picture per status | Onlies | .38 | .25 | .05 | 1790 | .96 |
| | Others | .38 | .21 | | | |
| picture rate | Onlies | .11 | .11 | .00 | 1790 | .93 |
| | Others | .11 | .14 | | | |
| **Social Habit** | | | | | | |
| **period for 1st status** | Onlies | 1.71 | 1.71 | -2.92 | 1790 | <.01 |
| | Others | 1.96 | 1.88 | | | |
| period for last status | Onlies | 5.92 | 0.42 | 1.13 | 1790 | .26 |
| | Others | 5.89 | 0.49 | | | |
| period for most statuses | Onlies | 5.20 | 1.48 | 0.17 | 1790 | .87 |
| | Others | 5.19 | 1.53 | | | |

From Table 3, it is obvious that there exists significant difference in 6 behaviors at level 0.01, 0.05 and 0.10 (the highlighted features in Table 3), and the six behaviors (the meanings of which have been described in Table 2) are:

description evaluation; mutual follower, follower; status, original picture;

period for 1st status, respectively.

From the perspective of "Personal Presentation", both Onlies and Others are willing to show themselves and their self-descriptions are generally optimistic. However, Onlies still differs a lot from Others significantly: Onlies usually describe themselves more positive than Others (2.24 vs 2.17, and *p<.05*). Moreover, Onlies are inclined to present a longer description, and they also label more tags for themselves in average. Generally speaking, Onlies are more willing to show themselves.

From another view of "Social Circle", we can easily distinguish Onlies from Others due to their different social circle. Onlies have significantly less mutual followers (130.65 vs 160.75, p<.01), also, they have significantly less followers (360.93 vs 371.43, p<.10), at the same time, they have less friends. Therefore, we can draw such a conclusion that Onlies' social circle is smaller.

In terms of "Social Activity", Onlies are found more willing to take part in social activities. Onlies post significantly more statuses (2834.19 vs 2642.42, p<.01), also, they upload significantly more original pictures (334.15 vs 236.58, p<.01). The other social activities are generally consistent with these two behaviors.

As for "Social Habit", our analysis indicates that Onlies have a different social habit from Others. Onlies usually post a day's first status earlier (1.71 vs 1.96, p<.01), and they post a day's last status later than Others.

**Significant Difference Analysis with Other Factors under Control**

We exploit independent sample t-test to check the effect of whether a subject is a single child on users' social media behaviors. However, what we discuss are just simple mean differences between single children and sibling children. The results could be due to differences in the demographic backgrounds of the selected samples. According to the previous research (Jiao, Ji, Jing, & Ching, 1986; Jiao, Ji, & Jing, 1996; Blake, 1981; Liu, Munakata, & Onuoha, 2005;

Cameron, Erkal, Gangadharan, & Meng, 2013), the following factors may also contribute to the behavioral difference of children: gender, age, growth area, culture extent and parents' culture extent.

To completely examine the relationship between whether a subject is a single child and social media behaviors, we take advantage of tobit regression model to estimate the true effect with other demographic factors under control. In other words, we want to explore the real effect of single child factor (SC). The regression equation pick up the possible influenced variables is listed as follows,

$$Y_i = \alpha + \beta X_i + \gamma Single_i + \varepsilon_i \quad (1)$$

Where $Y_i$ is a variable of the online behavior, and $X_i$ is a vector of control variables, which includes subjects' gender, growth area, culture extent, parents' culture extent and age; $Single_i$ is an indicator for being of an only child (denoted as SC), and it is also our main focus variable; $\varepsilon_i$ is a random term. The coefficients to be estimated are $\alpha$, $\beta$ and $\gamma$. Equation 1 can obviously check the real effect of being of a single child with other demographic variables under control. At first, we exploit Equation 1 to estimate the personal presentation of social media users. The analytical results are reported in Table 4.

Table 4

*Estimation of Demographical variables on personal presentation of social media users. The coefficients are identified in the table, and standard errors are followed in the bracket. \*, \*\*, \*\*\* indicate statistical significance at the 10,5 and 1% levels, respectively.*

| Personal Presentation | Screen name length | Description length | Description evaluation | Tag count |
|---|---|---|---|---|
| Single Child (SC) | 1.10(.39)*** | -2.66(3.00) | -.03(.04) | .67(.35)** |
| gender | -.45(.34) | -4.65(2.68)* | -.04(.03) | -.13(.33) |
| age | -.14(.03)*** | -.54(.25)** | .00(.00) | .03(.03) |
| Growth area | -.23(.18) | .22(1.40) | -.03(.02)* | .00(.00) |
| Culture extent | -.21(.13) | -4.37(1.05)*** | .01(.01) | -.07(.13) |
| Mother's culture extent | -.08(.15) | -.15(1.19) | -.05(.01)*** | -.13(.15) |
| Father's culture extent | .20(.14) | -.05(1.09) | 2.13(.09)*** | -.10(.14) |

Secondly, subjects' social circles are explored with the demographic variables. The analytical results are reported in Table 5.

Table 5

*Estimation of Demographical variables on social circle of social media users. The coefficients are identified in the table, and standard errors are followed in the bracket. *, **, *** indicate statistical significance at the 10,5 and 1% levels, respectively.*

| Social Circle | friend count | follower count | mutual follower count |
|---|---|---|---|
| Single Child (SC) | -6.93(18.01) | -469.35(258.15)* | -27.53(9.15)*** |
| gender | 78.20(16.08)*** | 731.19(247.93)*** | 60.62(8.17)*** |
| age | 11.05(1.49)*** | 43.85(22.79)* | 3.82(.76)*** |
| Growth area | -5.19(8.41) | .00(.00) | 2.12(4.27) |
| Culture extent | -2.95(6.31) | 33.88(97.14) | -2.52(3.21) |
| Mother's culture extent | 4.63(7.12) | 238.41(108.75)** | -1.25(3.62) |
| Father's culture extent | 3.10(6.56) | -153.10(100.94) | 1.20(3.33) |

Thirdly, we analyze the single child effect on social activities. We separate the analytical results in two tables: Table 6 and Table 7.

Table 6

*Estimation of Demographical variables on social activity of social media users-1. The coefficients are identified in the table, and standard errors are followed in the bracket. *, **, *** indicate statistical significance at the 10,5 and 1% levels, respectively.*

| Social Activity -1 | status | original status | Repost Status | favorite status | comment | Original picture |
|---|---|---|---|---|---|---|
| Single Child (SC) | 139.88( 74.21) ** | 26.17 (36.25) | -21.03 (22.47) | -21.29 (11.76) * | -212.36 (74.46) *** | -9.88 (12.05) |
| gender | 46.14 (66.25) | 175.26 (32.36) *** | 73.96 (20.06) *** | -51.17 (10.49) *** | 148.78 (66.45) ** | 39.92 (10.76) *** |
| age | 37.15 (6.13) *** | 9.68 (3.00) *** | 7.67 (1.86) *** | -2.75 (.97) *** | -15.45 6.14) ** | 7.30 (1.01) *** |
| Growth area | 28.76 34.63 | 3.06 (16.92) | 34.31 (10.48) *** | 4.73 (5.48) | 155.80 (34.72) *** | 11.18 (5.62) ** |
| Culture extent | -2.92 | 12.34 | -8.61 | 8.20 | 116.51 | -4.20 |

|  | 26.02 | (12.71) | 7.89 ** | (4.13) | (26.08) *** | (4.23) |
|---|---|---|---|---|---|---|
| Mother's culture extent | 11.21 29.31 | 3.97 (14.32) | 9.91 8.88 | 5.54 (4.65) | -15.53 (29.39) | 3.78 (4.76) |
| Father's culture extent | 17.19(27.02) | 9.45 (13.20) | 4.70 8.18 | 3.84 (4.28) | 27.38 (27.13) | 2.40 (4.39) |

Table 7

*Estimation of Demographical variables on social activities of social media users-2. The coefficients are identified in the table, and standard errors are followed in the bracket. *, **, *** indicate statistical significance at the 10,5 and 1% levels, respectively.*

| Social Activity -2 | original status rate | Comment variance | Original per status | Original pic rate |
|---|---|---|---|---|
| Single Child (SC) | -.01(.01) | -13.42(6.52)** | -.03(.01)*** | -.01(.01)** |
| gender | .05(.01)*** | 9.88(5.82)* | -.01(.01) | .02(.01)*** |
| age | -.00(.00) | -1.00(.54)* | .01(.00)\*** | .00(.00)*** |
| Growth area | -.00(.01) | 6.28(3.04)** | .02(.01)*** | .00(.00) |
| Culture extent | -.00(.00) | 1.43(2.29) | -.02(.00)*** | -.01(.00)*** |
| Mother's culture extent | -.00(.00) | -1.42(2.37) | .01(.00)** | .00(.00) |
| Father's culture extent | .00(.00) | 1.42(2.37) | .00(.00) | .00(.00) |

At last, we discuss the single child effect on social habit with the help of Equation 1. Table 8 reports the associated results.

Table 8

*Estimation of Demographical variables on social habits of social media users. The coefficients are identified in the table, and standard errors are followed in the bracket. \*, \*\*, \*\*\* indicate statistical significance at the 10,5 and 1% levels, respectively.*

| Social Habit | period for 1st status | period for last status | period for most statuses |
|---|---|---|---|
| Single Child (SC) | -.05(.17) | .09(.50) | -.65(.32)** |
| gender | -1.15(.16)*** | .78(.46)* | -.61(.28)** |
| age | .02(.01) | -.12(.04)*** | -.16(.02)*** |
| Growth area | -.01(.08) | .15(.23) | .22(.14) |
| Culture extent | -.22(.06)*** | .47(.16)*** | .48(.11)*** |
| Mother's culture extent | -.06(.07) | -.06(.20) | .01(.13) |
| Father's culture extent | -.01(.06) | .25(.18) | .15(.11) |

We find that even when the demographic and family background variables were controlled for the unconditional effects of single child we observed in Table 3 above still persist in terms of signs, magnitudes, and significance levels. Specifically speaking, the conclusions are consistent with unconditional analysis when other relative factors are under control, and the SC indicator indeed behave as a key factor to cause children's different behaviors on social media platform. The SC indicator generally perform as a significant factor to social media behaviors from Table 4 to Table 8.

## Discussion

Our results are a little different from previous researches that states no obvious difference exists between Onlies and Others, and is quite different from those take the generation of singleton as the "spoiled generation". Onlies even perform more active and positive on Weibo, and they are more willing to express themselves and communicate with others, though they have a smaller social circle.

In addition, Sina Weibo was founded in 2009, which is regarded as a novelty nowadays in China. Participants high on Openness would be expected to be active Weibo users and tend to perform multiple kinds of Weibo usage behaviors, such as posting more statuses and uploading more pictures. As to the Weibo descriptions, Onlies express more positive, which might because of more support from their parents. In our study, the average age of subjects is 23.3 (about 95% of subjects are birthed after 1979), and they are exactly the age to go into the society. Now in China, inflation, especially housing price, makes less trouble to those who can get support from their parents than others that cannot.

## Conclusion

In conclusion, we have investigated the difference between Onlies and Others, by comparing the personality traits and their Weibo behavior. We obtained 1792 copies of Sina Weibo data. We conducted independent t-test on the data to test if the only-child factors have effect on Weibo performances. Moreover, we take a further step to ensure the SC effect is not interfered from other factors. The data analysis show Onlies are generally more active and positive on Sina Weibo, while Onlies have a smaller social circle. We explained the phenomenon through one-child policy in China: one-child families are able to invest more to their children, and provide more support.

Apparently, there exists space we can do to put forward this exploration. In the future, we will check other factors which may affect the results by multi-way analysis. Meanwhile, we try to test more psychological factors, such as, mental health and social attitude. We want to know the effect of one-child policy on individual development.